\documentclass[twocolumn,10pt]{article}

\usepackage[a4paper,margin=2cm,footskip=6mm]{geometry}

\usepackage{booktabs} 
\usepackage{amsmath,amsthm,amssymb,amsfonts,mathtools}
\usepackage{algorithmic}
\usepackage{graphicx}
\usepackage{graphics}
\usepackage{textcomp}
\usepackage{xcolor}
\usepackage{layouts}
\usepackage{multirow}
\usepackage[ansinew]{inputenc}
\usepackage{mathtools}
\usepackage{tabularx}
\usepackage{multirow}
\usepackage{enumitem}
\usepackage{bm}
\usepackage[caption=false,font=footnotesize]{subfig}
\usepackage{mathtools}

\RequirePackage{fix-cm}

\DeclarePairedDelimiter\floor{\lfloor}{\rfloor}
\DeclarePairedDelimiter\ld{\text{ld(}}{)}

\newcommand{\cm}{\checkmark}
\newcommand{\ts}{\textsuperscript}
\newcommand\mydots{\makebox[1em][c]{...}}
\newcommand{\figwidthhalf}{1.57in}
\newcommand{\B}{\textbf}

\begin{document}

\date{}

\title{\Large \bf Accurate and Efficient Time Series Matching by Season- and Trend-aware Symbolic Approximation\\\vspace{3mm}
\large -- Extended Version Including Additional Evaluation and Proofs --
}

\author{
{\rm Lars Kegel, Claudio Hartmann, Maik Thiele, Wolfgang Lehner}\\
 TU Dresden, Database Systems Group\\
Dresden, Germany\\
\textless firstname.lastname\textgreater @tu-dresden.de
}

\maketitle

\thispagestyle{empty}

\subsection*{Abstract}
Processing and analyzing time series data\-sets have become a central issue in many domains requiring data management systems to support time series as a native data type.
A crucial prerequisite of these systems is time series matching, which still is a challenging problem.
A time series is a high-dimensional data type, its representation is storage-, and its comparison is time-consuming.
Among the representation techniques that tackle these challenges,
the symbolic aggregate approximation (SAX) is the current state of the art.
This technique reduces a time series to a low-dimensional space by segmenting it and discretizing each segment into a small symbolic alphabet.
However, SAX ignores the deterministic behavior of time series such as cyclical repeating patterns or trend component affecting all segments and leading to a distortion of the symbolic distribution.
In this paper, we present a season- and a trend-aware symbolic approximation.
We show that this improves the symbolic distribution and increase the representation accuracy without increasing its memory footprint.
Most importantly, this enables a more efficient time series matching by providing a match up to three orders of magnitude faster than SAX.\\
This is an extended version of \textit{Season- and Trend-aware Symbolic Approximation for Accurate and Efficient Time Series Matching} published in Datenbank Spektrum in 2021 \cite{Kegel2021}.

\section{Introduction}
Time series are the prime data source for data-mining tasks in many domains.
Exhaustive data gathering
and the specific characteristics of time series lead to the design of data management systems handling this data type natively \cite{Zoumpatianos2018}.
These systems provide a storage model, a query language, and optimization mechanisms suitable for time series.
Fast data access is needed to carry out complex data-mining tasks.
A crucial prerequisite of these systems is the retrieval of similar time series which is commonly referred to as \textit{time series matching} \cite{Zoumpatianos2018}.
Time series are high-dimensional due to their length which is why they are not directly matched against each other.
Instead, they are represented and compared in a low-dimensional space approximating their true (Euclidean) distance \cite{Agrawal1993b}.

For the past three decades, researchers have been developing representation techniques that rely on the shape \cite{Ding2008}, on model parameters \cite{Kalpakis2001}, or on the features \cite{Agrawal1993b} of a time series.
Among these techniques, the symbolic aggregate approximation (SAX) from Lin et al. is of particular interest \cite{Lin2003}.
First, this shape-based technique segments a time series into intervals which are represented by their mean value, this is called Piecewise Aggregate Approximation (PAA).
Second, it discretizes each mean value by mapping it to a discrete symbol.
Thus, SAX provides a small representation together with a fast distance measure which makes it suitable for time series matching.
Moreover, its distance measure has the important property to lower-bound the Euclidean distance measure, i.e.,
it allows for pruning observations based on the representation, without the need to load all high-dimensional time series into memory and calculate their Euclidean distance.

However, this technique suffers from two design criteria.
First, SAX assumes that the PAA of a normalized time series is also normally distributed with the same standard deviation.
As pointed out by Butler and Kazakov, this is over-simplistic and negatively impacts the symbolic distribution \cite{Butler2015}. 
Second, SAX ignores the deterministic behavior of a time series.
A \textit{season}, i.e., a cyclical repeated behavior, or a \textit{trend}, i.e., a long-term change in the mean level, are common deterministic components in many domains.
For example, production or consumption time series from the energy domain exhibit daily, weekly, or yearly seasons, while sales or price time series from the economy often exhibit an increasing or decreasing trend.
SAX does not take these features into account which again leads to a distortion in the symbolic distribution.

This work focuses on representation techniques based on SAX that aim to solve these two shortcomings and thus, provide a more efficient time series matching.

Several SAX extensions have been proposed to include mroe features.
While the original SAX focuses on the mean value of a segment,
existing extensions include standard deviation or trend.
To the best of our knowledge, no work in the literature focuses on global features before segmentation, i.e., features that arise from the deterministic behavior of a time series.

With this in mind, we introduce the season- and trend-aware symbolic approximations, \textit{sSAX} and \textit{tSAX}, that take a time series' season and trend into account.
First, they improve the symbolic distribution of the representation compared to SAX.
Second, they provide a more accurate representation while keeping the same representation size as SAX.
This enables these techniques to provide a more accurate and efficient time series matching.

In this paper, we make the following contributions:
\begin{itemize}
  \item We start with an overview on SAX and review existing extensions.
  While all extensions provide lower-bounding distance measures, most of them increase the representation size (Section \ref{sec:sota}).
  \item We introduce sSAX and tSAX which provide a higher matching accuracy at the same representation size as SAX.
  Their distance measures are lower-bounding, which is utmost important (Section \ref{sec:dSAX}). 
	\item Subsequently, we evaluate the techniques for time series matching.
	We summarize our experimental setting (Section \ref{sec:setting}), present, and discuss our results (Section \ref{sec:results}).
	The most remarkable result to emerge from this evaluation is that on large datasets (100 Gb), sSAX returns exact matches up to three orders of magnitude faster than SAX.
\end{itemize}
We conclude with future work in Section \ref{sec:concl_fw}.
\section{State of the Art}
\label{sec:sota}
We compare the original SAX with its extensions from the literature.
To do so, we begin with defining the terms time series dataset and time series matching along with its constraints from \cite{Lin2003}.
Second, we present the original SAX along with its representation and distance measure.
Third, we derive required properties that a representation shall support for efficient time series matching.
Finally, we review SAX extensions and compare them regarding these properties.

\subsection{Preliminaries}
\label{subsec:sota:prelim}
We adopt the following definition of a time series along with its constraints.
These constraints, which are also assumed by Lin et al. \cite{Lin2003}, are achieved by cleaning and transforming the data beforehand.

A \textit{time series} $\underline{x}$ is a vector of values $x$ which are measured at discrete time instances $t$:
\begin{equation}
\underline{x}^\intercal = (x_1, \mydots, x_t, \mydots, x_T) \text{~where~} \underline{x} \in \mathbb{R}^T, t\in \mathbb{N}_{>0}, t \leq T
\end{equation}
The time series is (1) finite with a fixed length $T$, (2) complete, i.e., without null values, and (3) equidistant, i.e., the distance between two time instances is constant.
Moreover, (4) it is normalized, i.e., its values have a sample mean of zero and a sample variance of one.

The goal of time series matching is to retrieve the most similar series out of a \textit{time series dataset} compared to a query time series.
A time series dataset is a set of $I$ time series with the same length:
\begin{equation}
\label{eq:sota:time_series_dataset}
	X = \{\underline{x}_1, \mydots, \underline{x}_i, \mydots, \underline{x}_I\} \text{~where~} i\in \mathbb{N}_{>0}, i \leq I
\end{equation}

The most similar observation has the lowest \textit{Euclidean distance} to the query:
\begin{equation}
\label{eq:sota:ed}
	d_{ED}(\underline{x}, \underline{x}') = \sqrt{\sum\nolimits_{t=1}^T (x_t - x'_t)^2}
\end{equation}
\newcommand*{\ed}{\ensuremath{d_{ED}(\underline{x}, \underline{x}')}}

Usually, time series are not directly compared using the Euclidean distance, since time series datasets may be quite large and calculating the Euclidean distances would require loading it into memory.
Moreover, the distance calculation is expensive as time series are high-dimensional.
These considerations motivate the representation of time series in a low-dimensional space.

\subsection{Original SAX}
The original SAX \cite{Lin2003} reduces the dimensionality of a time series in two steps:
First, the \textit{segmentation} into mean values by \textit{piecewise aggregate approximation} (PAA)  reduces the dimensionality in the time domain.
Second, the \textit{discretization} into symbols by \textit{symbolic aggregate approximation} (SAX) reduces the dimensionality in the value domain.
As such, PAA is a prerequisite of SAX, which is defined as follows.

Let $W\in\mathbb{N}_{>0}$ be the number of segments per time series, and $W$ divides $T$. 
The	PAA $\underline{\bar{x}}$ is the vector of mean values of a time series:
\begin{align}
\underline{\bar{x}}^\intercal &= (\bar{x}_1, \mydots, \bar{x}_w, \mydots, \bar{x}_W), \mbox{where}\\
\bar{x}_w &= \frac{W}{T} \sum_{t=\frac{T}{W}(w-1) + 1}^{\frac{T}{W}w}x_t
\label{eq:paa:mean} 
\end{align}

Although PAA reduces a time series in the time domain, it still contains real values that take a considerable amount of storage. Therefore, it is further reduced utilizing SAX.

Let $A$ be the size of an alphabet ($A\in\mathbb{N}_{>0}$) and let $\underline{b}^\intercal = (b_1, \mydots,$ 
$b_a, \mydots, b_{A-1})$ be a vector of increasingly sorted \textit{breakpoints} that split the real numbers into $A$ intervals:
\begin{equation}
]-\infty, b_1[, \mydots, [b_{a - 1}, b_{a}[, \mydots, [b_{A-1}, \infty[ 
\end{equation}
The SAX $\underline{\hat{x}}$ is the vector of symbols, i.e., mean values discretized into the alphabet $A$:
\begin{align}
\underline{\hat{x}}^\intercal &= (\hat{x}_1, \hat{x}_2, \mydots, \hat{x}_w, \mydots, \hat{x}_W), \mbox{where}\\
\hat{x}_w &= 
\begin{cases}
	1 & -\infty < \bar{x}_w < b_1\\
	a & \exists a:b_{a - 1} \leq \bar{x}_w < b_{a}\\
	A & b_{A - 1} \leq \bar{x}_w < \infty\\
\end{cases}
\end{align}

SAX reduces each mean value to a discrete symbol of the alphabet $A$.
It visualizes them with alphabetic characters (``a", ``b", \mydots) in order to stress their discrete nature.
Ideally, the symbols of a dataset are equiprobable so that they make full use of the alphabet capacity.
For achieving this, Lin et al. assume that mean values would be $\mathcal{N}(0,1)$-distributed because the time series are normalized \cite{Lin2003}.
Consequently, breakpoints are set such that the area under the normal distribution $\mathcal{N}(0, 1)$ from $[b_{a - 1}, b_{a}[$ equals $1/A$. 

Distance measures for PAA and SAX are defined as follows:
\begin{align}
d_{PAA}(\underline{\bar{x}}, \underline{\bar{x}}') &= \sqrt{^T/_W} \sqrt{\sum\nolimits_{w=1}^W(\bar{x}_w - \bar{x}'_w)^2}\\
d_{SAX}(\underline{\hat{x}}, \underline{\hat{x}}') &= \sqrt{^T/_W} \sqrt{\sum\nolimits_{w=1}^W cell(\hat{x}_w, \hat{x}'_w)^2}, \mbox{with}
\end{align}
\begin{align}
cell(a, a') &= 
\begin{cases}
	0 & |a - a'| \leq 1\\
	b_{max(a, a')} - b_{min(a, a')+1} & otherwise\\
\end{cases}
\label{eq:sota:cell}
\end{align}


\begin{figure}
	\centering
	\includegraphics[width=\linewidth]{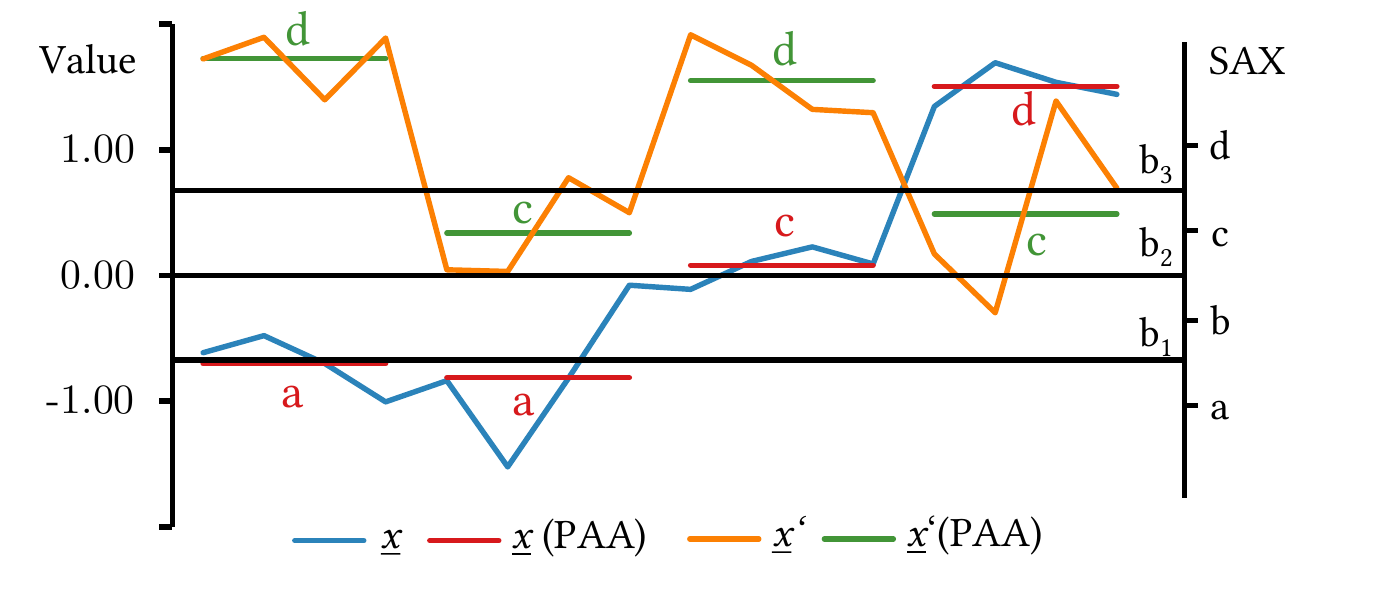}
	\caption{Time series with PAA and SAX representations}
	\label{fig:sax_repr}
\end{figure}

The example in Figure \ref{fig:sax_repr} shows a time series $\underline{x}$ (blue line, $T = 4$) from \cite{Shieh2008}.
Its PAA representation (red segments, $W = 4$) is $\underline{\bar{x}}^\intercal = (-0.70, -0.81, 0.08, 1.50)$.
Given an alphabet $A=4$ and respective breakpoints at $0.00$ and $\pm 0.67$ (black horizontal lines and x-axis), 
its SAX representation is $\underline{\hat{x}}^\intercal = (a, a, c, d)$.
 
The figure shows a second time series $\underline{x}'$ (orange line) whose PAA and SAX representations are 
$\underline{\bar{x}}'^\intercal = (1.72, 0.34,$ $1.55, 0.49)$ (green segments) and
$\underline{\hat{x}}'^\intercal = (d, c, d, c)$, respectively.
The Euclidean distance between $\underline{x}$ and $\underline{x}'$ is approx. 6.71, 
the PAA distance between $\underline{\bar{x}}$ and $\underline{\bar{x}}'$ is approx. 6.44, and 
the SAX distance between $\underline{\hat{x}}$ and $\underline{\hat{x}}'$ is approx. 3.02.

\subsection{Properties of Representation Technique}
Five properties of representation techniques are required for an efficient time series matching.
Sub\-sequently, we describe and review these properties regarding SAX.

\begin{description}[itemindent=0cm,leftmargin=0cm]
\item[\textbf{Representation Size}]
The representation size of a time series should be small compared to its original size.
Large time series datasets may not fit into memory and the matching may incur additional disk I/O.
Representation techniques are proposed to avoid these costs.
SAX reduces the time series to $W$ segments that are further reduced to symbols of an alphabet of size $A$.
This results in a representation size of $W\cdot \ld{A}$.
In Figure \ref{fig:sax_repr}, the SAX symbols of $\underline{\hat{x}}$ only need $4\cdot \ld{4} = 8~bits$.
This is small compared to the original time series $\underline{x}$ which needs $16\cdot 32 = 512~bits$, assuming that a real floating-point is stored with $32~bits$. 

\item[\textbf{Representation Time}]
The transformation of a time series into its representation should be fast.
SAX involves one pass over the data, carrying out segmentation and discretization simultaneously.
Thus, SAX allows for a fast transformation into a low-dimensional space.
The time series have to be already normalized.

\item[\textbf{Distance Storage}]
The distance calculation should not incur a substantial storage overhead.
SAX distance calculation involves the pairwise comparison of SAX symbols.
Instead of frequently recalculating these distances, Lin et al. store each symbol combination in a lookup table of size $A^2 \cdot 32~bits$ \cite{Lin2003}.
For a typical alphabet size of $A=256$, the lookup table's size is approx. $262~kb$, which is only calculated once for the dataset.

\item[\textbf{Distance Time}]
Time series matching can only benefit from a representation technique if the comparison between representations is faster than in the high-dimensional space.
The SAX distance calculation involves one lookup for each segment, so in total $W$ lookups for comparing two time series.
This is faster than the Euclidean distance where the calculation consists in loading two time series of length $T\gg W$ in memory and comparing them value by value.

\item[\textbf{Lower-bounding Distance}]
A distance measure is lo\-wer-bounding if the distance of two representations is always smaller than or equal to the true Euclidean distance of the original time series.
This property allows for pruning during time series matching: if the representation distance between a query and an observation is too large, there is no need to evaluate the Euclidean distance.
The PAA and SAX distance measures have been proven to lower-bound the Euclidean distance \cite{Yi2000,Lin2007}.
\end{description}

A representation technique that competes with SAX should provide similar properties.
Moreover, by including other features it should give a higher representation accuracy and a more efficient time series matching.

\subsection{SAX Extensions}
Several SAX extensions that include further features have been proposed in the literature.
We review their representation and distance properties and summarize them in Table \ref{tbl:rel_work:prop} together with the original SAX.
\begin{table}[t]
\renewcommand{\arraystretch}{1.3}
\caption{Properties of Representation Techniques}
\label{tbl:rel_work:prop}
\begin{minipage}{\columnwidth}
\centering
\resizebox{\columnwidth}{!}{%
\scriptsize
\begin{tabular}{c c c c c c}
\toprule
\multirow{2}{4.5em}{\bfseries Technique} 		& \multicolumn{2}{c} {\bfseries Representation} & \multicolumn{3}{c} {\bfseries Distance}\\
& \bfseries Size (bit)											& \bfseries Time	& \bfseries Storage (32 bits)	& \bfseries Time  & {\bfseries LB}\\
\midrule
SAX     & $W\cdot \ld{A}$               & 1 & $A^2$                     & $W$ & \cm \\
ESAX    & $3\cdot W\cdot \ld{A}$        & 1 & $A^6$                     & $W$ & \cm \\
1d-SAX  & $W\cdot \ld{A}$               & 2 &	$W\cdot A$          	&	$W$ & (\cm)\\
TD-SAX	& $W\cdot (\ld{A} + 32) + 32$   & 1 & $A^2$                     & $W$ & \cm \\
TFSA    & $W\cdot (\ld{T} + 66)$        & 3 & 0                         & $W$ & \cm \\
SAX\_SD & $W\cdot (\ld{A} + 32)$        & 1 & $A^2$	                    & $W$ & \cm \\
\hline
sSAX    & $W\cdot \ld{A}$               & 1 & $A_{seas}^2 + A_{res}^2$  & $4WL$   & \cm \\
tSAX	& $W\cdot \ld{A}$               & 2 & $A_{tr}^2 + A_{res}^2$    & $W + 1$    & \cm \\
\midrule
\end{tabular}%
}
\smallskip
{\scriptsize Representation Time (\#Passes over dataset); Distance Time (\#Lookups); LB: lower-bounding}
\end{minipage}
\end{table} 

ESAX \cite{Lkhagva2006} extends SAX by taking the extreme values of each segment into account.
Thus, it represents a PAA segment by minimum, mean, and maximum symbol, which are retrieved simultaneously.
Although this method is more accurate than SAX, it triples the representation size.
For each segment, the distance is calculated with the help of a lookup table.
However, the shape of the lookup table is not given.
Since there are three symbols per segment, a lookup table would have a size $A^6$.
Therefore, we assume that ESAX cannot fully precalculate the distances.

1d-SAX \cite{Malinowski2013} segments a time series first using piecewise linear approximation (PLA) and discretizes the values using SAX.
PLA represents each segment by its mean level and its slope which are estimated by linear regression.
Subsequently, these features are discretized like SAX and interleaved to one representation.
Thus, they have the same representation size as SAX.
For the representation, it needs a second pass over the data due to the linear regression.
The distance calculation is based on a lookup table and requires $W$ lookups.
It is only formulated for an asymmetric comparison, i.e., the distance of the real-valued query and the discretized observations.
Therefore, one lookup table is constructed per query that stores the distance between each segment of the query and each symbol of the alphabet.
   
TD-SAX \cite{Sun2014} also uses segment trends which are encoded as start and end value of each segment.
Although fast to calculate, these real values cause TD-SAX representations to be much larger than SAX.
The distance calculation for the mean values relies on $W$ lookups and thus, it needs the same storage as SAX.
However, the trend distance is calculated on real values and does not use a lookup table.
Therefore, distance calculation has additional costs. 

TFSA \cite{Yin2015} represents a time series by trends of segments only.
It splits the time series into segments of unequal length using a changepoint detection algorithm that passes twice over the time series.
On every segment, trends are extracted by linear regression (third pass) and discretized in 2 bits: increasing, decreasing or stationary.
The slope and the end point of each trend annotate this symbol, but they are not discretized, which increases representation size compared to SAX.
TFSA calculates the distance from the representation without a lookup table.
This practice might be slower because there are several floating-point operations involved for each segment.

SAX with standard deviation (SAX\_SD) \cite{Zan2016} represents every PAA segment with its mean value (discretized as SAX) and its standard deviation (not discretized).
Both features are calculated in one pass but they increase the representation size.
The distance calculation is done with a lookup table for the mean value.
The distance of the standard deviation is calculated directly on the feature.

Based on Table \ref{tbl:rel_work:prop}, we make the following observations: 
\begin{description}[itemindent=0cm,leftmargin=0cm]
  \item[\textbf{Representation Size}] All SAX extensions except 1d-SAX increase the representation size.
  For an unbiased evaluation of representation accuracy, it should be equal.
  \item[\textbf{Representation Time}] 1d-SAX and TFSA need several passes over the dataset for the representation, which is an acceptable penalty because the calculation still has linear complexity.
  \item[\textbf{Distance Storage}] The distance storage often uses a lookup table with a small size that provides a fast distance calculation.
  \item[\textbf{Distance Time}] The distance calculation needs $W$ look\-ups for all techniques. However, for features other than the mean value, the distance calculation has additional costs.
  A detailed evaluation would require an optimized distance function of each technique which is not covered in this work.
  \item[\textbf{Lower-bounding Distance}]
  The distance measures of ESAX, TD-SAX, TFSA, and SAX\_SD lower-bound the Euclidean distance measure.
  Although PLA is lower-bounding \cite{Chen2007}, it is not clearly stated for 1d-SAX.
\end{description}

Subsequently, we propose our symbolic approximations together with lower-bounding distance measures.
In contrast to the SAX extensions mentioned above, they provide a higher representation accuracy while having the same representation size as SAX, which is already represented in the lower part of Table \ref{tbl:rel_work:prop}.
\section{Season- and Trend-aware Symbolic Approximation}
\label{sec:dSAX}
Time series from many domains such as weather, energy, and economy, exhibit deterministic behavior.
The wind speed is often stronger in winter than in other seasons, while the solar irradiation has a strong daily season.
Consequently, this cyclical repeated behavior has an effect on the amount of renewable energy production.
In energy consumption, human behavior comes into play, where weekly patterns may be observed.
Economic time series may exhibit a trend due to increasing sales of some products.

Researchers have always applied SAX to datasets with trends and seasons, whether they were synthetic \cite{Lin2003,Lin2007} or real-world datasets \cite{Shieh2008}.
However, they did take this deterministic behavior into account and let it introduce redundancy into all symbols and distorted the symbolic distribution.
As a consequence, we argue that taking this behavior into account is essential for an accurate representation.

Therefore, we propose sSAX and tSAX that are aware of the time series' season and trend, respectively.
Each technique is described along with its time series model, representation, distance measure, and properties.
Moreover, both techniques take into account heuristics to improve the symbolic distribution.
These heuristics are efficiently calculated in the preprocessing step which is required by SAX to normalize the time series.

\subsection{Season-aware Symbolic Approximation - sSAX}
sSAX is aware of a time series' season by assuming there is a \textit{seasonal component} in the time series.
The remaining part of the time series forms the \textit{residuals} that are unstructured information.

\subsubsection{Time Series Model}
A season-aware time series model is shown in Eq. \eqref{eq:model_season}, where $\underline{seas}$ is the seasonal and $\underline{res}$ is the residual component of the time series $\underline{x}$:
\begin{equation}
\label{eq:model_season}
\underline{x} = \underline{seas} + \underline{res}
\end{equation}

A season repeats its behavior after $L$ values, this is called the \textit{season length}.
We adopt the additive season model, which is a common assumption in many domains and can also be used to represent multiplicative seasons.

The season is \textit{extracted} by averaging all values at the same seasonal position $l$ \cite{Kendall1983}.
The use of this technique, allows us to provide a fast representation and a lower-bounding distance measure.
A \textit{seasonal feature} $\sigma_l$ ($1\leq l\leq L$) is calculated as follows:
\begin{equation}
\label{eq:sigma}
\sigma_l = \frac{L}{T}\sum\nolimits_{k=1}^{T/L} x_{(k - 1) \cdot L + l} 
\end{equation}
where $T/L$ is the number of seasons in the time series that are iterated by $k$. 
The resulting features $\underline{\sigma}$ form the \textit{season mask}.

\begin{figure}[!t]
\centering
\subfloat[Time series]
{\includegraphics[width=\figwidthhalf]{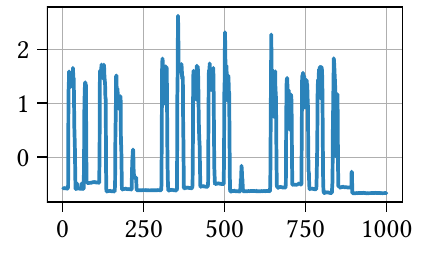}
\label{subfig:sSAX:seas_series}}
\hfil
\subfloat[Season mask] {\includegraphics[width=\figwidthhalf]{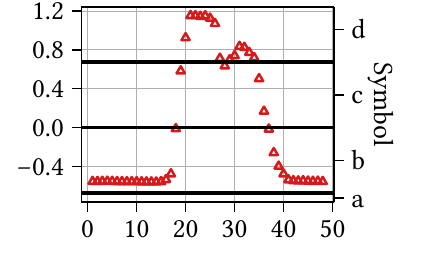}
\label{subfig:sSAX:seas_season}}
\caption{Time series with season}
\label{fig:sSAX}
\end{figure}

Figure \ref{fig:sSAX} illustrates a time series with a season that is repeated after 48 values (Figure \ref{subfig:sSAX:seas_series}).
Averaging the 1\ts{st}, 49\ts{th}, $\mydots$, and (T-L+1)\ts{th} value yields the 1\ts{st} of L seasonal features.
Averaging all positions results in the season mask (Figure \ref{subfig:sSAX:seas_season}).
This season mask is further discretized into an alphabet of size $A$, here $A = 4$.

\subsubsection{Representation}
Similar to SAX, the transformation in the low-dimen\-sional space is carried out in two steps: the season-aware PAA reduces the time series in the time domain, then the season-aware SAX reduces the value domain.

The \textit{season-aware PAA} (sPAA) combines the season mask and the PAA of the residuals in one representation.
While PAA would ignore the season of the time series by taking the mean value of a segment, sPAA explicitly extracts this season beforehand. 
Formally, the sPAA representation is the vector:
\begin{align}
	\underline{\bar{x}}^\intercal_{sPAA} = & (\sigma_1, \mydots, \sigma_l, \mydots, \sigma_L, \overline{res}_1, \mydots, \overline{res}_w, \mydots, \overline{res}_W), \nonumber\\
	&\mbox{where $W\cdot L$ divides $T$}.
\end{align}

This representation is made of real values and is further reduced by discretization of sSAX.
Let $A_{seas}, A_{res} \in \mathbb{N}_{>0}$ be the sizes of two alphabets.
Let $\underline{b}_{seas}$ and $\underline{b}_{res}$ be the respective vectors of breakpoints that split the real numbers into $A_{seas}$ and $A_{res}$ intervals.
Then, the sSAX representation is the vector $\underline{\hat{x}}_{sSAX}$:
\begin{equation}
\underline{\hat{x}}_{sSAX}^\intercal = (\hat{\sigma}_1, \mydots, \hat{\sigma}_l, \mydots, \hat{\sigma}_L, \widehat{res}_1, \mydots, \widehat{res}_w, \mydots, \widehat{res}_W)
\end{equation}
where $\hat{\sigma_l}$ is the symbol of $\sigma_l$ discretized into the alphabet $A_{seas}$ and $\widehat{res}_w$ is the symbol of $\overline{res}_w$ discretized into $A_{res}$.

The breakpoints are retrieved by two heuristics.
We quantify the influence of the season on the time series by the \textit{season strength} \cite{Wang2006}:
\begin{equation}\label{eq:seas_strength}
R^2_{seas} = 1 - \frac{var(\underline{res})}{var(\underline{x})}
\end{equation}
Assuming
(1) the season strength of the dataset is known,
(2) the time series are normalized, and
(3) the residual and seasonal component are independent of each other,
the following equations estimate the standard deviation of the season and the residuals:
\begin{align}
\label{eq:seas_sd_res}
sd(\underline{res}) &= \sqrt{1 - R^2_{seas}}\\
\label{eq:seas_sd_seas}
sd(\underline{seas}) &= \sqrt{1 - sd(\underline{res})^2}
\end{align}
where $R^2_{seas}$ is the mean season strength of the dataset.
Consequently, we set the breakpoints $\underline{b}_{seas}$ such that the area under normal distribution $\mathcal{N}(0, sd(\underline{seas}))$ is split into equiprobable regions $1/A_{seas}$.

Regarding the residuals, we also assume normally distributed mean values.
After season extraction the residual component has less influence and its variance does not achieve one as assumed from Lin et al.
Therefore, we set the breakpoints $\underline{b}_{res}$ such that the area under normal distribution $\mathcal{N}(0, sd(\underline{res}))$ is split into equiprobable regions $1/A_{res}$.

\subsubsection{Distance}
The distance measure of sPAA $d_{sPAA} (\underline{\bar{x}}_{sPAA}, \underline{\bar{x}}'_{sPAA})$ and sSAX $d_{sSAX}(\underline{\hat{x}}_{sSAX}, \underline{\hat{x}}'_{sSAX})$ are shown in Table \ref{tbl:dSAX:dist}.
sSAX relies on a lookup table that returns the precalculated distance of the season and residual symbols, using $\underline{b}_{seas}$ and $\underline{b}_{res}$ as breakpoints, respectively.
However, this lookup table for four symbols may get huge, which is why we propose an equivalent formulation for two smaller lookup tables.
Let $c_s$ be a lookup table defined as follows:
\begin{equation}
\label{eq:sseas:c_s}
c_s(a, a') = b_a - b_{a' + 1}
\end{equation}
where $\underline{b}_a$ are the breakpoints of the given feature.
Then \linebreak $cell(\hat{\sigma}, \hat{\sigma}', \widehat{res}, \widehat{res}')$ can be calculated by:
\begin{multline}
\label{eq:sseas:cell}
cell(\hat{\sigma}, \hat{\sigma}', \widehat{res}, \widehat{res}') = \\
\begin{cases}
	c_s(\hat{\sigma}, \hat{\sigma}') + c_s(\widehat{res}, \widehat{res}') & c_s(\hat{\sigma}, \hat{\sigma}') \geq -c_s(\widehat{res}, \widehat{res}') \\
	c_s(\hat{\sigma}', \hat{\sigma}) + c_s(\widehat{res}', \widehat{res}) & c_s(\hat{\sigma}', \hat{\sigma}) \geq -c_s(\widehat{res}', \widehat{res}) \\
	0 & otherwise
\end{cases}
\end{multline}

\newcolumntype{Y}{>{\centering\arraybackslash}X}
\begin{table*}[t]
\renewcommand{\arraystretch}{1.3}
\caption{Distance Measures}
\label{tbl:dSAX:dist}
\centering
\begin{tabularx}{0.97\textwidth}{Y c c}
\toprule
{\bfseries Technique}\xdef\tempwidth{\the\linewidth} & {\bfseries d\textsubscript{*PAA}} & {\bfseries d\textsubscript{*SAX}}\\
\midrule
\multicolumn{1}{m{\tempwidth}}{\centering sSAX}  										& $\sqrt{\frac{T}{W\cdot L}}\sqrt{\sum_{l=1}^L\sum_{w=1}^W (\sigma_l - \sigma'_l + \overline{res}_w - \overline{res}'_w)^2}$ & $\sqrt{\frac{T}{W\cdot L}}\sqrt{\sum_{l=1}^L\sum_{w=1}^W cell(\hat{\sigma}_l, \hat{\sigma}'_l, \widehat{res}_w, \widehat{res}'_w)^2}$\\
\\
\multicolumn{1}{m{\tempwidth}}{\centering tSAX}						& $\sqrt{\sum_{t=1}^T (\Delta\theta_1 + \Delta\theta_2\cdot (t - 1) + \Delta\overline{res}_{\lfloor(t-1)/(T/W)\rfloor+1})^2}$ & $\sqrt{c_t(\hat{\phi}, \hat{\phi}')^2 + \frac{T}{W} \sum_{w=1}^W cell(\widehat{res}_w, \widehat{res}'_w)^2}$\\
\bottomrule
\end{tabularx}
\end{table*}

\subsection{Trend-aware Symbolic Approximation - tSAX}
Similar to sSAX, tSAX is aware of the time series' trend and captures this behavior in a \textit{trend component}.

\subsubsection{Time Series Model}
A trend-aware time series model is shown in Eq. \eqref{eq:tr_tr}, where $\underline{tr}$ is the trend and $\underline{res}$ is the residual component of the time series $\underline{x}$. Again we adopt an additive combination:

\begin{equation}
\label{eq:tr_tr}
\underline{x} = \underline{tr} + \underline{res}
\end{equation}

Linear regression extracts the trend component from the time series.
This technique is fast and allows for proving the lower-bounding property of subsequent distance measures.
It estimates two features $\theta_1$ and $\theta_2$ which describe the \textit{base value} and the \textit{slope} of the time series, respectively.
Consequently, the trend-aware time series model is equal to:
\begin{equation}
\label{eq:tr_thetas}
\underline{x} = \theta_1 + \theta_2\cdot (\underline{t} - 1) + \underline{res}
\end{equation}
where $\underline{t}^\intercal=(1, \mydots, t, \mydots, T)$ is the vector of time instances of the time series.
Linear regression selects these features s.t. they minimize the sum of squared residuals $\sum_{t=1}^T res_t^2$.
Moreover, it yields two important properties.
First, the sum of the residuals is always zero, and
second, the trend component and the residuals are uncorrelated.

\begin{equation}
\label{eq:sum_res_0}
\sum\nolimits_{t=1}^T res_t = 0
\end{equation}
\begin{equation}
\label{eq:lr_uncorrel}
\sum\nolimits_{t=1}^T (tr_t \cdot res_t) = 0
\end{equation}

\begin{figure}[b]
\centering
\subfloat[Time series] {\includegraphics[width=\figwidthhalf]{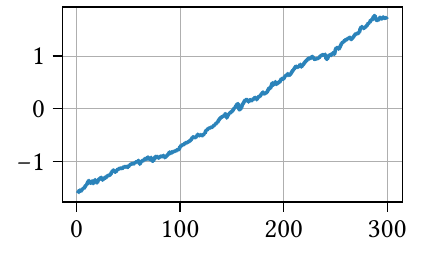}
\label{subfig:tSAX:tr_series}}
\hfil
\subfloat[Trend] {\includegraphics[width=\figwidthhalf]{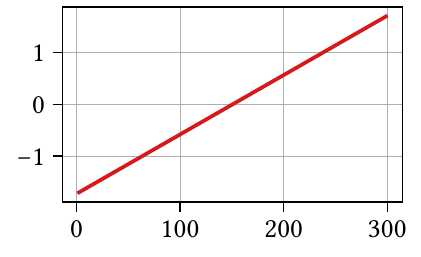}
\label{subfig:tSAX:tr_trend}}
\caption{Time series with trend}
\label{fig:tSAX}
\end{figure}

Figure \ref{fig:tSAX} represents a time series with a strong trend component (Figure \ref{subfig:tSAX:tr_series}).
Its trend is extracted by linear regression (Figure \ref{subfig:tSAX:tr_series}) and characterized by $\theta_1 = -1.72$ and $\theta_2 = 0.01$.

The features $\theta_1$ and $\theta_2$ are interdependent because the time series is normalized.
Therefore, the following equation holds:
\begin{equation}
\label{eq:tSAX:1f}
\theta_2 = -\frac{2}{T - 1}\cdot\theta_1
\end{equation}
The interested reader finds the proof in the Appendix (Subsection \ref{app:tSAX:1f}).
Using this equation, $\theta_1$ and $\theta_2$ are combined to one \textit{trend feature} $\phi$ that represents the angle between the x-axis and the trend component:
\begin{equation}
\phi = \arctan(\theta_2)  
\end{equation}

\subsubsection{Representation}
Similar to SAX, the trend-aware symbolic approximation transforms a time series in two steps:
first, the trend-aware PAA reduces the time series in the time domain;
second, the trend-aware SAX reduces it in the value domain.

The \textit{trend-aware PAA} (tPAA) representation is the vector of the trend feature and the mean values of a residual component:
\begin{equation}
\underline{\overline{x}}_{tPAA}^\intercal = (\phi, \overline{res}_1, \mydots, \overline{res}_w, \mydots, \overline{res}_W)
\end{equation}

Let $A_{tr}, A_{res} \in \mathbb{N}_{>0}$ be the sizes of two alphabets.
Let $\underline{b}_{tr}$ and $\underline{b}_{res}$ be the respective vectors of breakpoints that split the real numbers into $A_{tr}$ and $A_{res}$ intervals.
The \textit{trend-aware SAX} (tSAX) representation is the vector $\underline{\hat{x}}_{tSAX}$:
\begin{equation}
\underline{\hat{x}}_{tSAX}^\intercal = (\hat{\phi}, \widehat{res}_1, \mydots, \widehat{res}_w, \mydots, \widehat{res}_W)
\end{equation}
where $\hat{\phi}$ is the symbol of $\phi$ discretized into the alphabet $A_{tr}$ and $\widehat{res}_w$ is the symbol of $\overline{res}_w$ discretized into $A_{res}$.

Two heuristics retrieve the breakpoints.
As a result of normalization, there is a minimum and a maximum feature $\phi$ that is reached if the time series is a perfect trend with zero residuals.
Thus, $\phi$ is bounded by $\phi_{max}$: 
\begin{equation}
|\phi| \leq \phi_{max} \text{~where~} \phi_{max} = \tan^{-1} \sqrt{1/var(\underline{t})}
\end{equation}

Using this observation and the assumption that each trend is equiprobable, we set the breakpoints $\underline{b}_{tr}$ such that the area under uniform distribution between $[-\phi_{max}, \phi_{max}]$ is split into regions of probability $1/A_{tr}$.

Regarding the residuals, we adopt normally distributed mean values similar to SAX.
After extracting the trend, the residual component has less influence.
We quantify the influence of the trend on the time series by the \textit{trend strength} \cite{Wang2006}:
\begin{equation}\label{eq:tr_strength}
R^2_{tr} = 1 - \frac{var(\underline{res})}{var(\underline{x})}
\end{equation}
Assuming that 
(1) the trend strength of the dataset is known and
(2) the time series are normalized,
the standard deviation of the residuals is estimated by: 
\begin{equation}\label{eq:tr_sd_res}
sd(\underline{res}) = \sqrt{1 - R^2_{tr}}
\end{equation}
where $R^2_{tr}$ is the mean trend strength of the dataset.
Thus, we set $\underline{b}_{res}$ such that the area under normal distribution $\mathcal{N}(0, sd(\underline{res}))$ is split into equiprobable regions $1/A_{res}$.

\subsubsection{Distance}
tPAA and sPAA provide the distance measures $d_{tPAA}(\underline{\overline{x}}_{tPAA}, \underline{\overline{x}}'_{tPAA})$ and $d_{tSAX}(\underline{\hat{x}}_{tSAX}, \underline{\hat{x}}'_{tSAX})$ that are given in Table \ref{tbl:dSAX:dist}.
Let us note $\Delta f = f - f'$ as a shorthand symbol for the difference between a feature of time series $\underline{x}$ and $\underline{x}'$.
The tSAX distance measure relies on a lookup table $c_t$ for the trend feature using $\underline{b}_{tr}$ as breakpoints.
It expresses the minimum distance of two trend components represented by $\hat{\phi}$ and $\hat{\phi}'$.
For the residuals, tSAX relies on the lookup table $cell$ from SAX using $\underline{b}_{res}$ as breakpoints (Eq. \ref{eq:sota:cell}). 

\subsection{Properties of Representation Techniques}
We also review sSAX and tSAX regarding their properties (lower part of Table \ref{tbl:rel_work:prop}).
\begin{description}[itemindent=\parindent,leftmargin=0cm]
\item[Representation Size]
The alphabets $A_{seas}$, $A_{tr}$, and $A_{res}$ are chosen such that the representation size of sSAX and tSAX equals the representation size of SAX.
If they are not a power of 2, we allow for interleaving as in \cite{Malinowski2013}.
\item[Representation Time]
For sSAX, the representation needs one pass over the time series because season mask and residuals can be calculated simultaneously.
The tSAX representation needs an additional pass for the linear regression.
\item[Distance Storage]
Both representations need two look\-up tables of size $A_{res}^2$ and either $A_{seas}^2$ or $A_{tr}^2$.
Depending on the alphabet sizes, this leads to a storage for the distance calculation that is smaller or larger compared to SAX.
\item[Distance Time]
The sSAX distance measure needs at most $4\cdot W\cdot L$ lookups instead of $W$ lookups due to the combinations of season and residual symbols.
Although sSAX may use fewer segments for the residuals than SAX does for the time series, it leads to more lookups. 
However, if the calculation of the Euclidean distance is much slower if it incurs disk I/O, this limitation can be accepted.
The tSAX distance measure needs only $1$ lookup for the trend and $W$ lookups for the residuals.
\item[Lower-bounding Distance]
The most remarkable finding is that all presented distance measures, $d_{sPAA}$, $d_{tPAA}$, $d_{sSAX}$, and $d_{tSAX}$ lower-bound the Euclidean distance measure.
The interested reader finds the proofs in the Appendix (Subsections \ref{app:sPAA:lb}, \ref{app:sSAX:lb}, \ref{app:tPAA:lb}, and \ref{app:tSAX:lb}).
\end{description}
\section{Experimental Setting}
\label{sec:setting}
We compare our techniques sSAX and tSAX to the competitors SAX and 1d-SAX in order to examine the following hypotheses:
first, they improve the symbolic distribution of the residuals,
second, they provide a higher representation accuracy, and
third, they allow for a more accurate and efficient time series matching.

This section details the experimental setting, i.e.,
the matching methods that are applied,
the time series datasets that are selected,
the output variables that are measured,
the configurations chosen for the representation techniques, as well as 
the software and hardware environment.

\subsection{Matching Methods}
Besides the evaluation of the representation accuracy, it is common to assess a time series representation in a data-mining task.
In this work, it is applied to the \textit{exact} and \textit{approximate} time series matching.

Exact matching returns the observation from a time series dataset that has the minimum Euclidean distance to the query time series.
It conducts a \textit{linear search}:
First, the representation distance of the query to each observation in the dataset is calculated.
These distances are sorted increasingly.
Second, the Euclidean distance from the observations is calculated in the order of their representation distance, keeping track of the ``best-so-far'' observation and its Euclidean distance.
If this ``best-so-far'' (Euclidean) distance is less than the representation distance of the next observation, the linear search terminates and returns the ``best-so-far'' observation as an exact match.
The early termination is possible because of the lower-bounding property: subsequent observations never have a Euclidean distance that is smaller than the ``best-so-far'' distance.

Approximate matching returns the observation that is almost as good as the exact match.
It conducts a linear search on the representation, too.
However, it returns the observation with the minimum representation distance as an approximate match.
If there is more than one observation with the minimum representation distance, it returns the observation with the minimum Euclidean distance within this set.


\subsection{Datasets}
\label{subsec:setting:datasets}
To study the behavior of the representation techniques, they are evaluated on synthetic time series datasets with configurable characteristics and on two real-world time series datasets.
Table \ref{tbl:setting:dataset} recaps the dataset dimensions.
\begin{table}[t]
\caption{Dataset Dimensions}
\label{tbl:setting:dataset}
\centering
\small
\resizebox{\columnwidth}{!}{\begin{tabular}{c c c}
\toprule
Dataset				& Dataset Size $I$	& Length $T$ \\
\midrule
Season				 & 1,000							& [480; 960; 1,440; 1,920] \\
Trend					 & 1,000							& [480; 960; 1,440; 1,920] \\
Metering			 & 5,958							& 21840 \\
Economy				 & 6,400							& 300 \\
Season (Large) & [6,510,417; 13,020,833] & 960 \\
\bottomrule
\end{tabular}}
\normalsize
\end{table}

\begin{description}[itemindent=\parindent,leftmargin=0cm]
\item[Season]
A Season dataset contains 1,000 random walk time series, each of which is overlaid with a season mask of length 10.
In compliance with \cite{Shieh2008}, the time series length varies between 480 and 1,920.
All time series of a dataset have the same season strength which is fixed to a value between 1 and 99\%, where a tolerance of 0.5 percentage points (pp) in both directions is accepted.

\item[Trend]
A Trend dataset contains 1,000 random walk time series, each of which is overlaid with a trend.
Similarly to the Season datasets, the time series length varies between 480 and 1,920.
All time series of a dataset have the same trend strength between 1 and 99\% with a tolerance of 0.5 pp.

\item[Metering]
The Metering dataset is the result of the Smart Metering Project initiated from the Irish Commission for Energy Regulation \cite{Metering2015}. 
It contains the electricity consumption of households and small or medium businesses in Ireland between July 2009 and December 2010 measured in kilowatt-hour at a half-hour granularity.
All time series exhibit seasonal components but lack a strong trend.
The season-aware SAX is evaluated with respect to the daily season, which has, on average, a season strength of 18.3\%.

\item[Economy]
The Economy dataset contains about 100,000 time series from different domains (industry, finance, demographic, macro-/microeconomic, other).
It is the result from the M4-Compe\-tition to systematically evaluate the accuracy of forecast methods on a defined dataset \cite{Makridakis2018}. 
The values of each time series have a specified interval (year, quarter, month, other) and exhibit a trend component.
In compliance with the time series dataset (Eq. \ref{eq:sota:time_series_dataset}), only time series with the same length are selected, i.e., 6,400 time series measured for 25 years with monthly granularity.

\item[Season (Large)]
For the efficiency evaluation, two Season data\-sets with an overall size of 50 and 100 Gb are included.
The time series length is fixed to 960 values.
In contrast to Season, the season strength of a time series may vary.
We select datasets such that their season strength is on average 10.0\% (weak), 50.0\% (medium), and 90.0\% (strong).
\end{description}

For the accuracy evaluation, each time series from a dataset acts as query and is matched against the dataset of the remaining time series.
Thus, there are as many queries as there are time series in the dataset.
For the efficiency evaluation on Season (Large), we randomly select up to 50 query time series for each dataset.
We limit an experiment to four hours.
Since the runtime differs for each query, each technique is evaluated with the same set of queries.

\subsection{Output Variables}
Five output variables assess the accuracy and efficiency of symbolic approximations and enable us to evaluate our hypotheses:
the \textit{entropy}, the \textit{tightness of lower bound}, the \textit{pruning power}, the \textit{approximate accuracy}, and the \textit{runtime} which are defined subsequently.

\begin{description}[itemindent=\parindent,leftmargin=0cm]
\item[Entropy]
We hypothesize that sSAX and tSAX improve the distribution of the residual symbols compared to the SAX symbols.
A uniform distribution is a desirable property as Butler and Kazakov point out \cite{Butler2015}.
This property is quantified by the entropy as follows:
\begin{equation}
\label{eq:entropy}
H(A) = -\sum_{1\leq a\leq A} p(a)\cdot \ld{p(a)}
\end{equation}
where $p(a)$ is the frequency of a symbol in a SAX representation of a time series dataset.
An equal frequency of all symbols leads to the maximum entropy.
If some symbols are more frequent than others, the entropy decreases.
Only the entropy of two alphabets with equal size are compared since alphabets with different sizes have a different maximum entropy.

\item[Tightness of Lower Bound]
Our second hypothesis is that sSAX and tSAX provide a higher representation accuracy than the competitors.
This is evaluated with the tightness of lower bound (TLB) in accordance with \cite{Lin2003}.
The TLB expresses the ratio between the representation distance and the Euclidean distance as follows:  
\begin{equation}
\label{eq:tlb}
TLB(\underline{x}, \underline{x}') = \frac{d_{*SAX}(\underline{\hat{x}}, \underline{\hat{x}}')}{d_{ED}(\underline{x}, \underline{x}')}
\end{equation}
where $d_{*SAX}$ is either $d_{SAX}$, $d_{1d-SAX}$, $d_{sSAX}$, or $d_{tSAX}$.
To evaluate the TLB of a time series dataset, the mean TLB of all time series combinations is calculated.

\item[Pruning Power] Exact matching is improved by pruning observations to terminate linear search earlier. The pruning power (PP) expresses the fraction of observations that can be pruned without evaluating their Euclidean distance \cite{Chen2007}. A pruning power of $0$ means that no observations are pruned, a pruning power close to $1$ means that the linear search terminates after very few observation.

\item[Approximate Accuracy] Approximate matching is improved if the approximate match is closer to the Euclidean distance of the exact match. We introduce the output variable approximate accuracy (AA) which is the quotient of the Euclidean distance between the query and the exact match and the Euclidean distance between the query and the approximate match. An AA of $0$ means that the approximate match is very inaccurate, an AA of $1$ means that the approximate match is as accurate as the exact match.

\item[Runtime] The efficiency of time series matching is evaluated with the runtime. We measure the wall-clock time in seconds for the calculation of the representation distances and the Euclidean distances. 
\end{description}

\subsection{Configurations}
\label{subsec:setting:param}
Table \ref{tbl:setting:param} summarizes all possible configurations for the number of segments $W$ and the alphabet size $A$ for each dataset and each representation technique.
The representation size is fixed, which is why the alphabet $A_{res}$ of sSAX and of tSAX is set in accordance to $A_{seas}$ and $A_{tr}$, respectively.
1d-SAX uses the alphabet $A_a$ for the base value of a segment and the alphabet $A_s$ for the slope \cite{Malinowski2013}.
Alphabet sizes less or equal than 4 are ignored since they evidently cause a high accuracy loss.
We limit the size of a lookup table to $4~Mb$, which corresponds to an alphabet of size $1,024$. 
The standard deviation of sSAX and tSAX to discretize the residuals is derived from the component strength (Eqs. \ref{eq:seas_sd_res} and \ref{eq:tr_sd_res}). 

\begin{table}[t]
\renewcommand{\arraystretch}{1.3}
\caption{Configurations of Representation Techniques}
\label{tbl:setting:param}
\resizebox{\columnwidth}{!}{%
\begin{tabular}{c c c c c}
\toprule
 {\bfseries Synthetic} & {\bfseries W} & {\bfseries A or A\textsubscript{res}} & {\bfseries A\textsubscript{seas} or A\textsubscript{tr}} & {\bfseries Size / bit}\\
 \midrule
SAX & [32; 40; 48; 96] & [1,024; 256; 101; 10] & - & 320\\
sSAX & [24; 48; 48] & [1,024; 32; 64] & [256; 256; 9] & 320\\
tSAX & [32; 40; 48; 96] & $\floor*{2\hat{\text{~}}((320-\ld{A_{tr}})/W)}$ & [32; 128; 1,024] & 320\\
\midrule
{\bfseries Metering} & {\bfseries W} & {\bfseries A or A\textsubscript{res}} & {\bfseries A\textsubscript{seas} or A\textsubscript{tr}} & {\bfseries Size / bit}\\
\midrule
SAX & [455; 520; 728; 910] & [256; 128; 32; 16] & - & 3,640\\
sSAX & 455 & [191; 165; 142; 123] & [16; 64; 256; 1024] & 3,640\\
\midrule
{\bfseries Economy} & {\bfseries W} & {\bfseries A, A\textsubscript{a}, or A\textsubscript{res}} & {\bfseries A\textsubscript{tr} or A\textsubscript{s}} & {\bfseries Size / bit}\\
\midrule
SAX & [10; 12; 15; 20; 30] & [256; 101; 40; 16; 6] & - & 80\\
1d-SAX & [10; 12; 15; 20] & $\floor*{2\hat{\text{~}}((80-\ld{A_s}\cdot W)/W)}$ & [8; 16; 32] & 80\\
tSAX & [10; 12; 15; 20; 30] & $\floor*{2\hat{\text{~}}((80-\ld{A_{tr}})/W)}$ & [16; 64; 256; 1,024] & 80\\
\bottomrule
\end{tabular}%
}
\end{table}

\subsection{Software and Hardware Environment}
All representation techniques are implemented using R \cite{R2018}.
For the runtime evaluation, matching methods are implemented in C and compiled with GCC 6.3.0 with level 3 optimization under Windows 10.
Experiments run on a machine with Intel(R) i7 Processor 6660U@2.60GHz and 20 Gb of RAM and compare two disks, one 1 Tb HDD and one 500 Gb SSD.  
Each time series is stored as a binary file on disk.
Time series representations and lookup tables are kept in-memory, while time series are read from disk without system cache buffering.
\section{Results and Discussion}
\label{sec:results}
The results of our evaluation are presented and discussed in the order of our hypotheses.
The quality of the symbolic distribution is assessed,
followed by the evaluation of the representation accuracy.
Finally, the accuracy and efficiency of matching methods are evaluated.

\subsection{Symbolic Distribution}
We compare the symbolic distribution of SAX, sSAX, and tSAX utilizing the entropy.
For this evaluation, the alphabet size is the same for all configurations and is fixed to $A=A_{res}=256$ which results in a maximum entropy of $H_{max}(A)=8$.
\begin{figure}[!b]
\centering
\subfloat[Season by series length] {\includegraphics[width=\figwidthhalf]{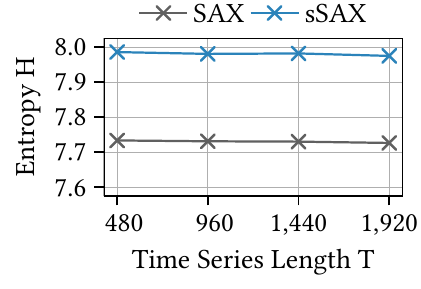}
\label{subfig:entr_seas:T}}
\hfil
\setcounter{subfigure}{3}%
\subfloat[Trend by series length] {\includegraphics[width=\figwidthhalf]{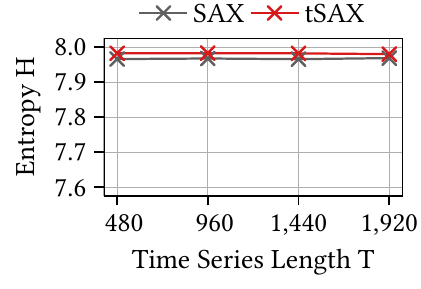}
\label{subfig:entr_tr:T}}
\\
\setcounter{subfigure}{1}%
\subfloat[Season by number of segment] {\includegraphics[width=\figwidthhalf]{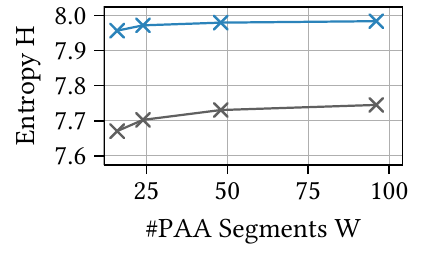}
\label{subfig:entr_seas:w}}
\hfil
\setcounter{subfigure}{4}%
\subfloat[Trend by number of segment] {\includegraphics[width=\figwidthhalf]{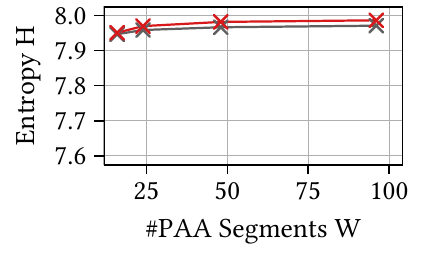}
\label{subfig:entr_tr:w}}
\\
\setcounter{subfigure}{2}%
\subfloat[Season by strength] {\includegraphics[width=\figwidthhalf]{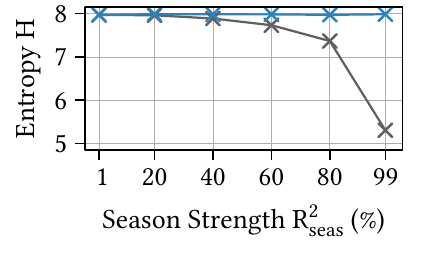}
\label{subfig:entr_seas:R^2_seas}}
\hfil
\setcounter{subfigure}{5}%
\subfloat[Trend by strength] {\includegraphics[width=\figwidthhalf]{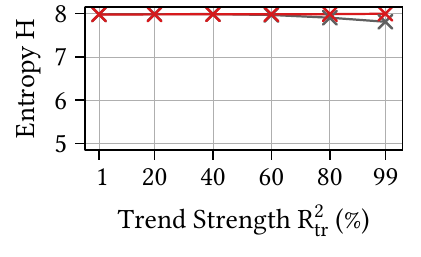}
\label{subfig:entr_tr:R^2_tr}}
\caption{Entropy on synthetic datasets}
\label{fig:entr}
\end{figure}

Figure \ref{fig:entr} visualizes the entropy of SAX and sSAX on the Season datasets.
The sSAX technique systematically provides a higher entropy, which results in more equally distributed symbols.
The entropy decreases if time series get longer (Figure \ref{subfig:entr_seas:T}) or if there are fewer segments, i.e., longer segments (Figure \ref{subfig:entr_seas:w}).
This observation confirms the observation of Butler and Kazakov that the mean value distorts the symbolic distribution \cite{Butler2015}.
However, this effect is less strong for sSAX.
Interestingly, the entropy of SAX significantly decreases for datasets with a strong season, but it is not the case for sSAX (Figure \ref{subfig:entr_seas:R^2_seas}).
This finding confirms that the seasonal component should be treated separately.

Figure \ref{fig:entr} also visualizes the entropy of SAX and tSAX on the Trend datasets.
Figures \ref{subfig:entr_tr:T} and \ref{subfig:entr_tr:w} reconfirm the finding that the mean value distorts the symbolic distribution and that tSAX remediates this effect.
Moreover, Figure \ref{subfig:entr_tr:R^2_tr} shows that the entropy decreases if SAX is applied to a time series with strong trends and that this entropy is more stable for tSAX.
Overall, both techniques exhibit a high entropy close to the maximum.
Therefore, the gain of tSAX over SAX is less strong than the gain of sSAX over SAX.

On the real-world dataset, we compare the techniques for the same number of windows $W$.
On the Metering dataset, the entropy increases from 6.96 for SAX to 7.09 for sSAX.
On the Economy dataset, the entropy increases from 7.92 for SAX to 7.95 for tSAX.
The overall increase of entropy is less strong compared to the homogeneous synthetic datasets.
This result is due to the heterogeneous component strengths in both datasets, and due to the heuristics which assume the mean strength.

\subsection{Representation Accuracy}
\begin{figure}[!b]
  \centering
  \subfloat[Season]
  {\includegraphics[width=\figwidthhalf]
  {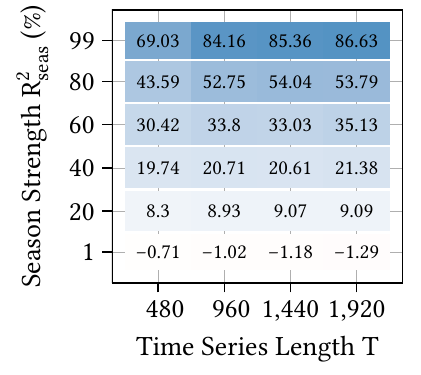}
  \label{subfig:lb:season}}
  \hfil
  \subfloat[Trend] {\includegraphics[width=\figwidthhalf]
  {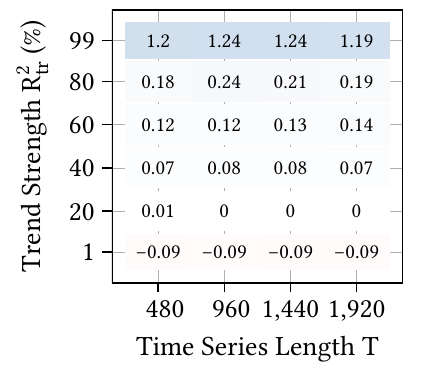}
  \label{subfig:lb:trend}}
  \\
  \subfloat[Metering] {\includegraphics[width=\figwidthhalf]
  {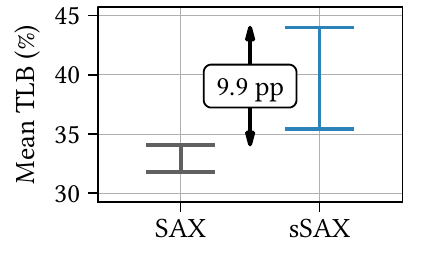}
  \label{subfig:lb:metering}}
  \hfil
  \subfloat[Economy]
  {\includegraphics[width=\figwidthhalf]
  {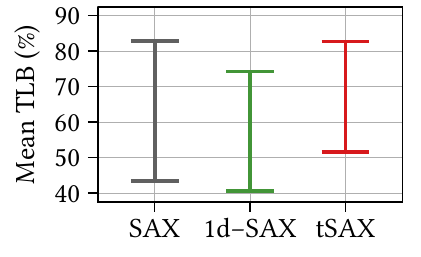}
  \label{subfig:lb:economy}}
  \caption{Increase of TLB compared to SAX}
  \label{fig:lb}
\end{figure}

We evaluate the representation accuracy utilizing the TLB (Figure \ref{fig:lb}). 
For this evaluation, the representation size is constant for each dataset, and the possible configurations are given in Table \ref{tbl:setting:param}.

On the synthetic datasets, the TLB of SAX is compared to sSAX and tSAX (Figures \ref{subfig:lb:season} and \ref{subfig:lb:trend}).
Results are grouped by time series length and component strength.
Each cell presents the difference in percentage points between the mean TLB of the most accurate sSAX/tSAX configuration and the mean TLB of the most accurate SAX configuration.
Figure \ref{subfig:lb:season} shows that sSAX gains accuracy compared to SAX with up to 86 pp.
The longer the time series and the stronger the season, the higher is the accuracy gain.
If there is no season, sSAX is only slightly less accurate.
The tSAX technique gains accuracy by only 1.2 pp and has slight losses in the absence of a trend (Figure \ref{subfig:lb:trend}).
This gain is lower than expected.
However, there are two possible reasons for this.
First, the distance of discretized mean values from SAX satisfactorily captures the global trend of a time series. 
Second, the normalization transforms the time series such that the change in the mean level becomes smaller.
Therefore, tSAX has not much room for improvement. 

Figures \ref{subfig:lb:metering} and \ref{subfig:lb:economy} display the results for the real-world datasets. It shows the minimum and maximum mean TLB which are reached with the chosen configurations.
On the Metering dataset, the best sSAX configuration gains up to 9.9 pp compared to the best SAX configuration (Figure \ref{subfig:lb:metering}).
Thus, if the season is taken into account, it leads to a much higher representation accuracy.
On the Economy dataset, we include 1d-SAX in our comparison, which is the only trend-aware SAX extension that has the same representation size as SAX.
Overall, tSAX has a better representation accuracy than 1d-SAX.
Thus, it better takes advantage of the available representation size.
However, it does not gain compared to the best SAX configurations and reaches at best 82.7\% while SAX reaches 82.9\% (Figure \ref{subfig:lb:economy}). 

\subsection{Exact Matching}
\begin{figure}[!b]
  \centering
  \subfloat[Season]
  {\includegraphics[width=\figwidthhalf]
  {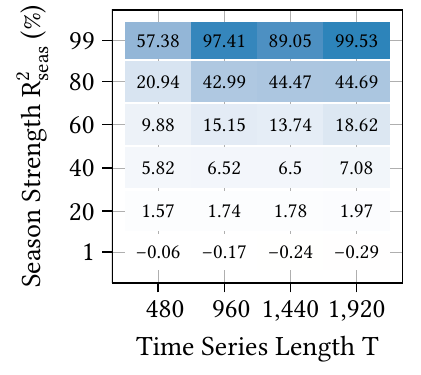}
  \label{subfig:pp:season}}
  \hfil
  \subfloat[Trend] {\includegraphics[width=\figwidthhalf]
  {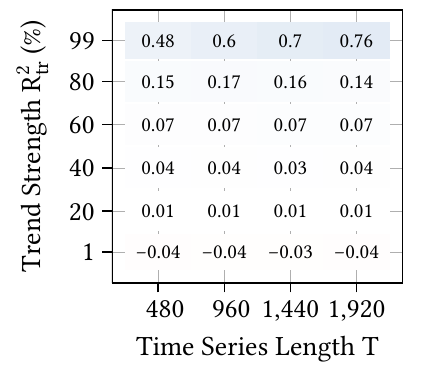}
  \label{subfig:pp:trend}}
  \\
  \subfloat[Metering] {\includegraphics[width=\figwidthhalf]
  {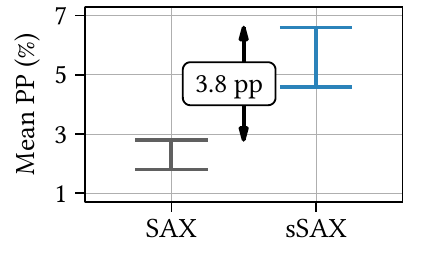}
  \label{subfig:pp:metering}}
  \hfil
  \subfloat[Economy]
  {\includegraphics[width=\figwidthhalf]
  {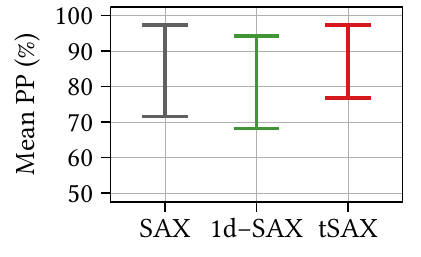}
  \label{subfig:pp:economy}}
  \caption{Increase of pruning power compared to SAX}
  \label{fig:pp}
\end{figure}

For exact matching, we first evaluate the pruning power that results from the representation accuracy (Figure \ref{fig:pp}).

On the synthetic datasets, sSAX and tSAX exhibit a gain in pruning power compared to SAX.
Remarkably, sSAX improves the pruning power up to 99 pp in the presence of a strong season (Figure \ref{subfig:pp:season}).
However, if no season is present, sSAX has a worse pruning power by at most 0.29 pp which is negligible regarding the overall gain of sSAX.
The tSAX technique improves the pruning power even for weak trends (Figure \ref{subfig:pp:trend}).
But the gain is limited with at most 0.76 pp.

This behavior is also confirmed on the real-world datasets.
On Metering, sSAX gains 3.8 pp in pruning power and reaches 6.6\% (Figure \ref{subfig:pp:metering}).
While SAX can prune 274 from 5,958 time series on average, sSAX efficiently prunes 393 time series.
On Economy, the best SAX configuration already has a very high pruning power (Figure \ref{subfig:pp:economy}) with 97.4\%.
The tSAX technique outperforms 1d-SAX but reaches at best the same pruning power as SAX.
\newcommand{\sshead}[1]{\multicolumn{2}{c} {$\bm{R^2_{seas} = #1\%} $ }}
\begin{table}[t]
\renewcommand{\arraystretch}{1.3}
\caption{Matching Efficiency on Season (Large)}
\label{tbl:eff}
\begin{minipage}{\columnwidth}
\centering
\resizebox{\columnwidth}{!}{%
\begin{tabular}{c c c c c c c c c}
\toprule
\B{HDD} 												&															&															& \sshead{10.0}														& \sshead{50.0}													& \sshead{90.0}\\
\B{Size}												& \B{Technique}								& \bfseries Repr. 						& \bfseries Raw 		& \bfseries Sum			 	& \bfseries Raw & \bfseries Sum 			& \bfseries Raw	& \bfseries Sum			 	\\
\midrule
\multirow[c]{2}{*}{\B{50 Gb}}		& SAX 												& 1.80												& 135.82						& 137.61							& 1,801.12			& 1,802.92							& 6046.75				& 6048.55						\\			
																& sSAX 												& 8.67												& 41.76 						& \B{50.43}					 	& 0.54					& \B{9.21}							& 0.08					& \B{8.75}					\\
\midrule
\multirow[c]{2}{*}{\B{100 Gb}}	& SAX													& 3.69												& 73.61							&	77.30								& 4181.02				& 4184.72								& 13,423.47			& 13,427.16					\\
																& sSAX												& 16.86												& 4.77							& \B{21.63}						& 1.09					& \B{17.95}							& 0.11					& \B{16.97}					\\
\midrule
\B{SSD} 												&															&															& \sshead{10.0}														& \sshead{50.0}													& \sshead{90.0}\\
\B{Size}												& \B{Technique}								& \bfseries Repr. 						& \bfseries Raw 		& \bfseries Sum			 	& \bfseries Raw & \bfseries Sum 				& \bfseries Raw	& \bfseries Sum			\\
\midrule
\multirow[c]{2}{*}{\B{50 Gb}}		& SAX 												& 1.84												& 4.05							& \B{5.89}						& 101.61				&	103.45								& 850.81					& 852.65					\\
																& sSAX 												& 9.12												& 0.71							& 9.83								&	0.04					&	\B{9.16}							& 0.02						& \B{9.14}				\\
\hline
\multirow[c]{2}{*}{\B{100 Gb}}	& SAX													& 3.80												& 8.29							& \B{12.09}						& 115.14				& 118.95								& 1,088.80				& 1,092.60				\\
																& sSAX												& 17.99												& 1.05							& 19.04								& 0.07					& \B{18.06}							& 0.02						& \B{18.02}				\\
\bottomrule
\end{tabular}%
}
\smallskip
{\scriptsize Repr.: Mean Runtime for Representation Distance Calculation;\\
\vspace{-0.15cm}
Raw: Mean Runtime for Euclidean Distance Calculation}
\end{minipage}
\end{table}

Let us now look at the efficiency evaluation. 
Table \ref{tbl:eff} details
the runtimes for both disks (HDD, SSD) on the Season (large) datasets with 50Gb and 100Gb.
The runtime is broken down into one part for calculating the representation distances including result ordering (Repr.),
and in another part for accessing the time series and calculating the true distances (Raw).
For each season strength, the sum of both parts indicates the mean runtime per query.
Calculating the representation distances is displayed once for all season strengths since it does not depend on this heuristic.

The tables reveal that
(1) sSAX is faster for all datasets from HDD even when there is only a weak season strength,
(2) sSAX is faster for all datasets from SSD for a significant season strength.
The most striking result to emerge from the data is that sSAX is up to three orders of magnitude faster for time series with a strong season.
On HDD, sSAX requires approximately 17 seconds for querying the 100 Gb dataset, while SAX requires approximately 3.7 hours. 
SAX has a decreased pruning power and thus, needs more disk access.
The pruning power of sSAX, however, increases and provides exact matches even faster.
A na\"ive matching of a query without representation technique would require 6,137 seconds (50 Gb on SSD) and 13,866 seconds (100 Gb on SSD); thus it is much slower than sSAX.
We did not evaluated na\"ive matching on HDD due to time restrictions.

\subsection{Approximate Matching}
\begin{figure}[!t]
  \centering
  \subfloat[Season]
  {\includegraphics[width=\figwidthhalf]
  {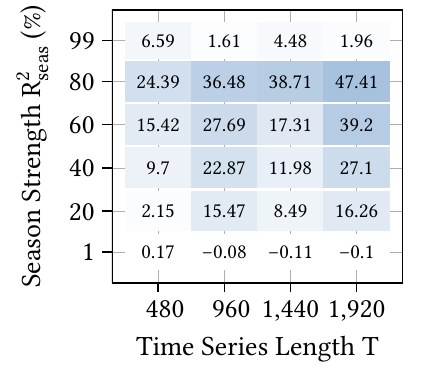}
  \label{subfig:aa:season}}
  \hfil
  \subfloat[Trend] {\includegraphics[width=\figwidthhalf]
  {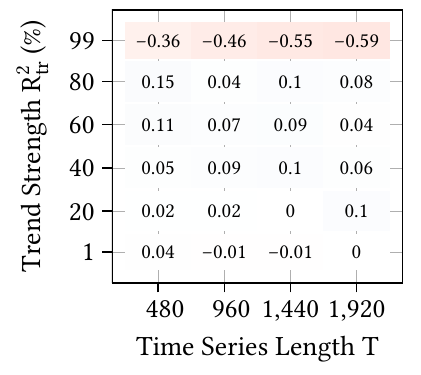}
  \label{subfig:aa:trend}}
  \\
  \subfloat[Metering]
  {\includegraphics[width=\figwidthhalf]
  {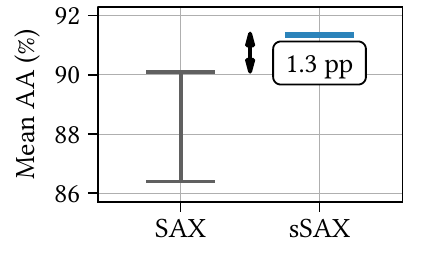}
  \label{subfig:aa:metering}}
  \hfil
  \subfloat[Economy] {\includegraphics[width=\figwidthhalf]
  {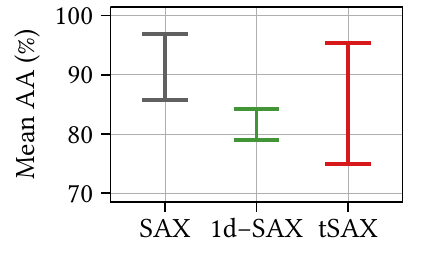}
  \label{subfig:aa:economy}}
  \caption{Increase of approximate accuracy compared to SAX}
  \label{fig:aa}
\end{figure}

In approximate matching, we evaluate the accuracy of an approximate match compared to the exact match utilizing the approximate accuracy (Figure \ref{fig:aa}).

Figures \ref{subfig:aa:season} and \ref{subfig:aa:trend} show the increase of approximate accuracy on the synthetic datasets by time series length and component strength.
The longer the time series and the stronger the deterministic component, the higher is the gain of sSAX and tSAX over SAX.
The sSAX reaches up to 47 pp improvement on Season, i.e., the approximate match is, on average, 50\% more accurate (Figure \ref{subfig:aa:season}).
Due to the aforementioned reasons, tSAX only reaches minor improvements on Trend with up to 0.14 pp (Figure \ref{subfig:aa:trend}).
On datasets with a strong component strength, both representations reach an approximate accuracy of approximately 99\%.
The sSAX and tSAX representations reach this accuracy thanks to the accurate representation.
SAX reaches this accuracy because most of the observations have the same representation which is why SAX re-evaluates their Euclidean distance to the query in order to retrieve the most accurate approximate observation.
Thus, it reaches a slightly higher approximate accuracy, which is mainly due to the evaluation of the Euclidean distance.

On the real-world datasets, sSAX and tSAX show a similar behavior.
All sSAX configurations outperform all SAX configurations on Metering, and the best sSAX configuration is up to 1.3 pp more accurate than the best SAX configuration (Figure \ref{subfig:aa:metering}).
The exact search shows that the time series are all very close to each other regarding the Euclidean distance.
The pruning power of 6.6\% shows that for many observations, it has to be checked whether they are the exact match (Figure \ref{subfig:pp:metering}).
However, the approximate match already reaches approximately 91.5\% of the accuracy of the exact match.
Although tSAX provides a higher approximate accuracy compared to 1d-SAX (95.4\% vs. 84.2\%), it cannot reach the best approximate accuracy of SAX (96.9\%) (Figure \ref{subfig:aa:economy}).

Let us now look at the efficiency evaluation.
For all chosen queries, there are not two matches with the same minimum representation distance (which would require the calculation of the Euclidean distance).
Therefore, we can focus on the representation distance calculation.
Table \ref{tbl:eff} reveals that approximate matching with sSAX is slower compared to SAX due to an increased number of lookups.
However, these approximate matches are much more accurate, as Figure \ref{fig:aa} suggests.

\section{Conclusion and Future Work}
\label{sec:concl_fw}



We have proposed two novel symbolic approximations sSAX and tSAX to improve representation accuracy and time series matching compared to state-of-the-art techniques. Our evaluation shows that considering the deterministic features during time series matching is worth the effort. It improves the representation accuracy, especially if the deterministic component is strong, and the accuracy of exact as well as approximate matching. Moreover, exact matching with sSAX is more efficient by up to three orders of magnitude. Even if exact matching was carried out with indexes based on SAX such as iSAX and its successors \cite{Shieh2008,Zhang2019}, it could not avoid the disk access for Euclidean distance calculation. While sSAX provides significant improvements, the improvements of tSAX are interestingly far less significant.

In the future, we will concentrate on representing combinations of deterministic components since time series usually exhibit (several) seasonal components simultaneously in combination with an (potentially non-linear) trend.

Furthermore, recent work has focused on indexes based on SAX for matching billions of time series \cite{Zoumpatianos2016,Zhang2019}.
However, both works analyzed rather short time series ($T \leq 640$) and our approximations have the potential to efficiently index and match much longer time series thanks to their higher representation accuracy.

\bibliographystyle{plain}


\appendix
\section{Appendix}
\subsection{Proof of Lower-bounding sPAA}
\label{app:sPAA:lb}

\newcommand*{\fracf}{\ensuremath{\frac{T}{W\cdot L}}}
\newcommand*{\sumT}{\ensuremath{\sum_{t=1}^T}}
\newcommand*{\sumE}{\ensuremath{\sum_{j=1}^E}}
\newcommand*{\sumL}{\ensuremath{\sum_{j=1}^L}}
\newcommand*{\sumB}{\ensuremath{\sum_{k=1}^B}}
\newcommand*{\sumBL}{\ensuremath{\sum_{\substack{1\leq k\leq B \\ 1\leq j\leq L}}}}
\newcommand*{\sigmat}{\ensuremath{\sigma_{(t-1)\%L+1}}}
\newcommand*{\respaat}{\ensuremath{\overline{res}_{\lfloor\frac{t-1}{E}\rfloor+1}}}

\begin{proof}
We show that the following inequality always holds:
\begin{equation}
d_{sPAA} (\underline{\bar{x}}_{sPAA}, \underline{\bar{x}}'_{sPAA})
\leq d_{ED}(\underline{x}, \underline{x}')
\end{equation}
For convenience, we use $E=T/W$ for the length of a segment and $B=T/L$ for the number of seasons in the time series.
We square both sides:
\begin{equation}
\label{eq:sSAX:lb:squaring}
d_{sPAA} (\underline{\bar{x}}_{sPAA}, \underline{\bar{x}}'_{sPAA})^2
\leq \ed^2
\end{equation}
Subsequently and until Eq. \ref{eq:sSAX:lb:as1234}, we focus on transforming the left-hand side.
Using $d_{sPAA}$ from Section \ref{sec:dSAX} yields:
\begin{equation}
\label{eq:sSAX:lb:formula}
\fracf\sumT (\Delta\sigmat + \Delta\respaat)^2
\end{equation}
Rewriting the residuals with Eq. \ref{eq:paa:mean} from PAA yields:
\begin{equation}
\label{eq:sSAX:lb:formulaRewrite}
\fracf\sumT(\Delta\sigmat + \frac{1}{E}\sumE\Delta res_{\lfloor\frac{t-1}{E}\rfloor\cdot E + j})^2
\end{equation}
The residuals are the difference of time series values and season mask:
\begin{align}
\label{eq:sSAX:lb:diffRawSeason}
\fracf\sumT\big( & + \Delta\sigmat \notag\\
& + \frac{1}{E}\sumE\Delta x_{\lfloor\frac{t-1}{E}\rfloor\cdot E + j}\notag\\
& - \frac{1}{E}\sumE\Delta\sigma_{(\lfloor\frac{t-1}{E}\rfloor\cdot E + j - 1)\%L+1}\big)^2
\end{align}
The season mask is also expressed as mean values of $\underline{x}$ (Eq. \ref{eq:sigma}):
\begin{align}
\label{eq:sSAX:lb:applySeasonMask}
\fracf\sumT\big(& + \frac{1}{B}\sum_{k=1}^{B} \Delta x_{(k - 1) \cdot L + (t-1)\%L+1}\notag\\
& + \frac{1}{E}\sumE\Delta x_{\lfloor\frac{t-1}{E}\rfloor\cdot E + j}\notag\\
& - \frac{1}{B}\frac{1}{E}\sumE\sum_{k=1}^{B} \Delta x_{(k - 1) \cdot L + (\lfloor\frac{t-1}{E}\rfloor\cdot E + j - 1)\%L+1}\big)^2
\end{align}
We focus on the case when $W\cdot L = T$, thus $E=L$ and $W=B$.
Factoring out $\frac{1}{B\cdot L}=\frac{1}{T}$ yields:
\begin{align}
\label{eq:sSAX:lb:case1}
\frac{1}{T^2}\sumT\big( & + L\cdot\sum_{k=1}^{B} \Delta x_{(k - 1) \cdot L + (t-1)\%L+1}\notag \\
& + B\cdot\sumL\Delta x_{\lfloor\frac{t-1}{L}\rfloor\cdot L + j} \notag\\
& - \sumL\sum_{k=1}^{B} \Delta x_{(k - 1) \cdot L + j}\big)^2
\end{align}
We arrange the distances $\Delta x_t$ in a matrix $D\in \mathbb{R}^{B\times L}$ as follows:
\begin{equation}
D =
 \begin{pmatrix}
  \Delta x_1 &   \Delta x_2  &   \ldots & \Delta x_L \\
  \Delta x_{L+1} & \Delta x_{L+2} & \ldots & \Delta x_{2\cdot L} \\
  \vdots & \vdots & \ddots & \vdots \\
  \Delta x_{(B-1)\cdot L+1} & \Delta x_{(B-1)\cdot L+2} & \ldots & \Delta x_{T} \\
 \end{pmatrix}
\end{equation}
Consequently, we arrange the factors $L$ and $B$ and $-1$ in a matrix $A^t\in  \mathbb{R}^{B\times L}$.
The positions of these factors depend on $t$.
The cell of t which is in row  $\lfloor\frac{t-1}{L}\rfloor+1$ and column $(t-1)\%L+1$ is filled with $B + L - 1$.
The other cells in the same row are filled with $B-1$.
The other cells in the same column are filled with $L-1$.
All other cells are filled with $-1$.
For example, for $t = 1$:
\begin{equation}
A^1 =
 \begin{pmatrix}
B + L - 1	&	B - 1	& \ldots 	& B - 1\\
L - 1		&	-1	& \ldots 	& -1\\
\vdots 	& \vdots 	& \ddots 	& \vdots \\
L - 1		&	-1	& \ldots 	& -1\\
 \end{pmatrix}
\end{equation}
With these two matrices, Eq. \ref{eq:sSAX:lb:case1} is represented as follows:
\begin{equation}
\label{eq:sSAX:lb:asMatrix}
\frac{1}{T^2}\sumT\big(\sumB\sumL A^t_{j,k} D_{j,k}\big)^2
\end{equation}
Expanding the square and iterating through $t$ yields: 
\begin{align}
\label{eq:sSAX:lb:as1234}
\frac{1}{T} \big(& + (B + L -1 ) \sumBL D^2_{k,j}\notag\\ 
& + (B - 1) \sumBL \sum_{\substack{1\leq j'\leq L \\ j'\neq j}} D_{k,j} \cdot D_{k,j'}\notag\\
& + (L - 1) \sumBL \sum_{\substack{1\leq k'\leq B \\ k'\neq k}} D_{k,j} \cdot D_{k',j}\notag\\
& - \sumBL\sum_{\substack{1\leq k'\leq B \\ 1\leq j'\leq L \\ k'\neq k \\ j'\neq j}} D_{k,j} \cdot D_{k',j'}\big)
\end{align}
The Euclidean distance from the right-hand side (Eq. \ref{eq:sSAX:lb:squaring}) can be rewritten as $\ed^2 = \sumBL D_{k, j}^2$ and can be subtracted from the left-hand side.
Multiplying by $-T$ leads to:
\begin{align}
\label{eq:sSAX:lb:as1234MinusED}
& + (B\cdot L + 1 - B - L)\sumBL D^2_{k,j}\\
& - (1 - B) \sumBL \sum_{\substack{1\leq j'\leq L \\ j'\neq j}} D_{k,j} \cdot D_{k,j'}\\
& -(1 - L) \sumBL \sum_{\substack{1\leq k'\leq B \\ k'\neq k}} D_{k,j} \cdot D_{k',j}\\
& + \sumBL\sum_{\substack{1\leq k'\leq B \\ 1\leq j'\leq L \\ k'\neq k \\ j'\neq j}} D_{k,j} \cdot D_{k',j'} \geq 0
\end{align}
This is equivalent to:
\begin{equation}
\label{eq:sSAX:lb:final}
\underset{k<k'}{\sum^B_{k=1}\sum^B_{k'=1}}\text{~}\underset{j<j'}{\sum^L_{j=1}\sum^L_{j'=1}} (D_{k, j} - D_{k, j'} - D_{k', j} + D_{k', j'})^2 \geq 0
\end{equation}
Since all summands are squared, they are always greater than or equal to zero.
The other cases $W\cdot L \neq T$ where $W\cdot L$ divides $T$ can be solved similarily.
\end{proof}

\subsection{Proof Lower-bounding sSAX}
\label{app:sSAX:lb}
\begin{proof}
We show that sSAX lower-bounds the Euclidean distance:
\begin{equation}
d_{sSAX}(\underline{\hat{x}}_{sSAX}, \underline{\hat{x}}'_{sSAX}) \leq d_{ED}(\underline{x}, \underline{x}')
\end{equation}
We show this property based on the distance measure from Eq. \ref{eq:sseas:cell}.
For given symbols $\hat{\sigma}, \hat{\sigma}', \widehat{res}, \widehat{res}'$, we assume that the Case 1 from Eq. \ref{eq:sseas:cell} holds:
$c_s(\hat{\sigma}, \hat{\sigma}') \geq -c_s(\widehat{res}, \widehat{res}')$.
Using Eq. \ref{eq:sseas:c_s} and reformulating the inequality taking into account the sPAA features yields:
\begin{equation}
\sigma + \overline{res} \geq
b_{\hat{\sigma}} + b_{\widehat{res}} \geq 
b_{\hat{\sigma}' + 1} + b_{\widehat{res}' + 1} \geq
\sigma' + \overline{res}'
\end{equation}
Thus, $cell(\hat{\sigma}, \hat{\sigma}', \widehat{res}, \widehat{res}')$ is always lower than the sum of sPAA features:
\begin{align}
\MoveEqLeft[3] cell(\hat{\sigma}, \hat{\sigma}', \widehat{res}, \widehat{res}')\notag\\
={}& c_s(\hat{\sigma}, \hat{\sigma}') + c_s(\widehat{res}, \widehat{res}')\notag\\
={}& b_{\hat{\sigma}} - b_{\hat{\sigma}' + 1} + b_{\widehat{res}} - b_{\widehat{res}' + 1} \notag\\
={}& b_{\hat{\sigma}} + b_{\widehat{res}} - (b_{\hat{\sigma}' + 1} + b_{\widehat{res}' + 1}) \notag\\
\leq{}& \sigma + \overline{res} - (\sigma' + \overline{res}')
\end{align}
By symmetry, this inequality also holds for Case 2 of Eq. \ref{eq:sseas:cell} and it is trivial for Case 3.
Thus, we can conclude:
\begin{equation}
d_{sSAX}(\underline{\hat{x}}_{sSAX}, \underline{\hat{x}}'_{sSAX})
\leq d_{sPAA}(\underline{\bar{x}}_{sPAA}, \underline{\bar{x}}'_{sPAA})
\end{equation}
As shown in Subsection \ref{app:sPAA:lb}, sPAA distance lower-bounds the Euclidean distance.
\end{proof}

\subsection{Proof of Eq. \ref{eq:tSAX:1f} (Combined trend feature)}
\label{app:tSAX:1f}
\begin{proof}
As defined in Subsection \ref{subsec:sota:prelim}, a time series is normalized.
Thus, the sample mean of its values is zero.
Taking into account the trend component (Eqs. \ref{eq:tr_tr} and \ref{eq:tr_thetas}):
\begin{equation}
\sum_{t=1}^T x_t = \sum_{t=1}^T(tr_t + res_t) = \sum_{t=1}^T(\theta_1 + \theta_2\cdot (t - 1) + res_t) = 0
\end{equation}
Linear regression implies that the residuals' sum is zero (Eq. \ref{eq:sum_res_0}):
\begin{equation}
\sum_{t=1}^T(\theta_1 + \theta_2\cdot (t - 1)) = T\cdot (\theta_1 + \theta_2\cdot \frac{T - 1}{2}) = 0
\end{equation}
\end{proof}

\subsection{Proof of Lower-bounding tPAA}
\label{app:tPAA:lb}
\begin{proof}
We show that tPAA lower-bounds the Euclidean distance:
\begin{equation}
d_{tPAA}(\underline{\overline{x}}_{tPAA}, \underline{\overline{x}}'_{tPAA}) \leq d_{ED}(\underline{x}, \underline{x}')
\end{equation}
For convenience, we rewrite $\overline{res}_t = \overline{res}_{\lfloor(t-1)/(T/W)\rfloor+1}$.
Squaring each side and setting the trend component yields:
\begin{equation}
\sum_{t=1}^T (\Delta tr_t + \Delta \overline{res}_t)^2
\leq
\sum_{t=1}^T (\Delta tr_t + \Delta res_t)^2
\end{equation}
Expanding the summands and subtracting the common summands:
\begin{align}
\sum_{t=1}^T \big((\Delta\overline{res}_t)^2 + 2\cdot \Delta tr_t\cdot\Delta \overline{res}_t\big)
\leq \nonumber \\
\sum_{t=1}^T \big((\Delta res_t)^2 + 2\cdot \Delta tr_t\cdot\Delta res_t\big)
\end{align}
All summands except one are arranged on the right-hand side:
\begin{equation}
\sum_{t=1}^T (\Delta \overline{res}_t)^2
\leq 
\sum_{t=1}^T \big(
	(\Delta res_t)^2
	+ 2\cdot \Delta tr_t \cdot (\Delta res_t - \Delta \overline{res}_t)
\big)
\end{equation}
Trend components and residuals are not correlated which is also true for mean residuals (Eq. \ref{eq:lr_uncorrel}).
Thus, this equation can be rewritten as:
\begin{equation}
\sum_{t=1}^T (\Delta \overline{res}_t)^2 \leq \sum_{t=1}^T (\Delta res_t)^2
\end{equation}
PAA distance lower-bounds the Euclidean distance as shown in \cite{Yi2000}.
\end{proof}

\subsection{Proof of Lower-bounding tSAX}
\label{app:tSAX:lb}
\begin{proof}
We show that tSAX lower-bounds the Euclidean distance:
\begin{equation}
d_{tSAX}(\underline{\hat{x}}_{tSAX}, \underline{\hat{x}}'_{tSAX}) \leq d_{ED}(\underline{x}, \underline{x}')
\end{equation}
By construction, the lookup table $c_t$ always returns the minimum distance between two trend components represented by $\hat\phi$ and by $\hat\phi'$.
Similarily, the lookup table $cell$ returns the minimum distance between two PAA mean values $\widehat{res}_w$ and $\widehat{res}'_w$.
Therefore:
\begin{align}
\MoveEqLeft[3] d_{tSAX}(\underline{\hat{x}}_{tSAX}, \underline{\hat{x}}'_{tSAX}) \\
={}& \sqrt{c_t(\hat\phi, \hat\phi')^2 + \frac{T}{W} \sum\nolimits_{w=1}^W cell(\widehat{res}_w, \widehat{res}'_w)^2} \\
\leq{}& \sqrt{\sum\nolimits_{t=1}^T (\Delta^2 tr_t + \Delta^2 \overline{res}_t)} \\
={}&  \sqrt{\sum\nolimits_{t=1}^T \big((\Delta tr_t)^2 + (\Delta \overline{res}_t)^2 + 2\cdot \Delta tr_t\cdot \Delta \overline{res}_t\big)} \label{eq:tSAX:lb:mix}\\
={}&  \sqrt{\sum\nolimits_{t=1}^T (\Delta tr_t + \Delta \overline{res}_t)^2}\\
={}&  d_{tPAA}(\underline{\overline{x}}_{tPAA}, \underline{\overline{x}}'_{tPAA})
\end{align}
In Eq. \ref{eq:tSAX:lb:mix}, we introduce a summand that equals zero because trend and residuals are not correlated (Eq. \ref{eq:lr_uncorrel}).
As already shown, tPAA distance lower-bounds the Euclidean distance (Appendix \ref{app:tPAA:lb}). 
\end{proof}

\end{document}